\documentclass[prl,aps,twocolumn,superscriptaddress,showpacs,showkeys,amsmath,amssymb,floatfix]{revtex4-1}
\usepackage[colorlinks=true,citecolor=blue,filecolor=blue,linkcolor=blue,urlcolor=blue,pdftex]{hyperref}

\usepackage{xspace} 
\usepackage[usenames,dvipsnames]{color}
\usepackage{booktabs,graphicx,mathrsfs,verbatim,amsmath,units,soul}
\usepackage{xspace} 
\usepackage{upgreek}
\usepackage{gensymb}
\usepackage{amsmath}
\usepackage{tabularx}


\newcommand{\bologna}{\affiliation{Department of Physics and Astronomy, University of Bologna and INFN-Bologna, 40126 Bologna, Italy}}

\newcommand{\chicago}{\affiliation{Department of Physics \& Kavli Institute for Cosmological Physics, University of Chicago, Chicago, IL 60637, USA}}

\newcommand{\coimbra}{\affiliation{LIBPhys, Department of Physics, University of Coimbra, 3004-516 Coimbra, Portugal}}

\newcommand{\columbia}{\affiliation{Physics Department, Columbia University, New York, NY 10027, USA}}

\newcommand{\lngs}{\affiliation{INFN-Laboratori Nazionali del Gran Sasso and Gran Sasso Science Institute, 67100 L'Aquila, Italy}}

\newcommand{\mainz}{\affiliation{Institut f\"ur Physik \& Exzellenzcluster PRISMA, Johannes Gutenberg-Universit\"at Mainz, 55099 Mainz, Germany}}

\newcommand{\heidelberg}{\affiliation{Max-Planck-Institut f\"ur Kernphysik, 69117 Heidelberg, Germany}}

\newcommand{\munster}{\affiliation{Institut f\"ur Kernphysik, Westf\"alische Wilhelms-Universit\"at M\"unster, 48149 M\"unster, Germany}}

\newcommand{\nikhef}{\affiliation{Nikhef and the University of Amsterdam, Science Park, 1098XG Amsterdam, Netherlands}}

\newcommand{\nyuad}{\affiliation{New York University Abu Dhabi, Abu Dhabi, United Arab Emirates}}

\newcommand{\purdue}{\affiliation{Department of Physics and Astronomy, Purdue University, West Lafayette, IN 47907, USA}}

\newcommand{\rpi}{\affiliation{Department of Physics, Applied Physics and Astronomy, Rensselaer Polytechnic Institute, Troy, NY 12180, USA}}

\newcommand{\rice}{\affiliation{Department of Physics and Astronomy, Rice University, Houston, TX 77005, USA}}

\newcommand{\stockholm}{\affiliation{Oskar Klein Centre, Department of Physics, Stockholm University, AlbaNova, Stockholm SE-10691, Sweden}}

\newcommand{\subatech}{\affiliation{SUBATECH, IMT Atlantique, CNRS/IN2P3, Universit\'e de Nantes, Nantes 44307, France}}

\newcommand{\torino}{\affiliation{INAF-Astrophysical Observatory of Torino, Department of Physics, University  of  Torino  and  INFN-Torino,  10125  Torino,  Italy}}

\newcommand{\ucla}{\affiliation{Physics \& Astronomy Department, University of California, Los Angeles, CA 90095, USA}}

\newcommand{\ucsd}{\affiliation{Department of Physics, University of California, San Diego, CA 92093, USA}}

\newcommand{\wis}{\affiliation{Department of Particle Physics and Astrophysics, Weizmann Institute of Science, Rehovot 7610001, Israel}}

\newcommand{\zurich}{\affiliation{Physik-Institut, University of Zurich, 8057  Zurich, Switzerland}}

\newcommand{\paris}{\affiliation{LPNHE, Universit\'{e} Pierre et Marie Curie, Universit\'{e} Paris Diderot, CNRS/IN2P3, Paris 75252, France}}

\newcommand{\freiburg}{\affiliation{Physikalisches Institut, Universit\"at Freiburg, 79104 Freiburg, Germany}}

\newcommand{\lal}{\affiliation{LAL, Universit\'e Paris-Sud, CNRS/IN2P3, Universit\'e Paris-Saclay, F-91405 Orsay, France}}

\newcommand{\naples}{\affiliation{Department of Physics ``Ettore Pancini'', University of Napoli and INFN-Napoli, 80126 Napoli, Italy}} 

\newcommand{\nagoya}{\affiliation{Kobayashi-Maskawa Institute for the Origin of Particles and the Universe, Nagoya University, Furo-cho, Chikusa-ku, Nagoya, Aichi 464-8602, Japan}}

\begin{document}
%\linenumbers
% CHANGE TO TITLE
%\title{Search for light WIMP induced electronic recoils through Migdal effect in XENON1T}
\title{A Search for Light Dark Matter Interactions Enhanced by the Migdal effect or Bremsstrahlung in XENON1T}

\author{E.~Aprile}\columbia
\author{J.~Aalbers}\stockholm
\author{F.~Agostini}\bologna
\author{M.~Alfonsi}\mainz
\author{L.~Althueser}\munster
\author{F.~D.~Amaro}\coimbra
\author{V.~C.~Antochi}\stockholm
\author{E.~Angelino}\torino
\author{F.~Arneodo}\nyuad
\author{D.~Barge}\stockholm
% ...

\collaboration{XENON Collaboration}
\email[]{xenon@lngs.infn.it}
\noaffiliation

\date{\today} 

\begin{abstract}
%XENON1T experiment has reported the results with 1\,tonne $\times$ year exposure for the search of Weakly Interacting Massive Particles (WIMPs), by looking for the nuclear recoils induced by elastic scatters between WIMPs and xenon nuclei.
%The results give the world-leading constraint of WIMP with mass larger than 6\,GeV.
%Based on the same data, in this letter we present a search for sub-GeV WIMPs by probing the electronic recoils that are induced by radiations in the inelastic channel of WIMP-xenon nuclear recoils through Bremsstrahlung or Migdal effect.
Direct dark matter detection experiments based on a liquid xenon target are leading the search for dark matter particles with masses above $\sim$\,5\,GeV/c$^2$, but have limited sensitivity to lighter masses because of the small momentum transfer in dark matter-nucleus elastic scattering.
% It has been recently suggested that DM-nucleus elastic scattering can lead to electronic recoil signals induced by the Bremsstrahlung and Migdal effect, which are eliminated as background events in the previous searches. 
%However in the atomic view, the DM-nucleus scattering occurs inelastically and leads to a polarization of the recoiling atom between nucleus and atomic shell electrons.
However, there is an irreducible contribution from inelastic processes accompanying the elastic scattering, which leads to the excitation and ionization of the recoiling atom (the Migdal effect) or the emission of a Bremsstrahlung photon.
% These new detection channels enable to lower the detector threshold of LXe detector and significantly enhances the sensitivity to sub-GeV DM. 
In this letter, we report on a probe of low-mass dark matter with masses down to about 85\,MeV/c$^2$ by looking for electronic recoils induced by the Migdal effect and Bremsstrahlung, using data from the XENON1T experiment. 
%We exploit an approach that detects ionization signal only, as well as both scintillation and ionization signals, which enables to lower the detection threshold.
Besides the approach of detecting both scintillation and ionization signals, we exploit an approach that uses ionization signals only, which allows for a lower detection threshold.
This analysis significantly enhances the sensitivity of XENON1T to light dark matter previously beyond its reach.

% The most stringent limits on the spin-independent WIMP-nucleon interaction cross-section of 3.6$\times$10$^{-36}$\,cm$^2$ and 3.1$\times$10$^{-39}$\,cm$^2$ for WIMP masses of 0.1 GeV/c$^2$ and 1 GeV/c$^2$, respectively, are obtained at 90\% confidence level.
%we derive the first limits on DM-nucleon scattering for mχ < 500 MeV
\end{abstract}



\maketitle

%%%%%%%%%%%%%%%%%%%%%%%%%
% Introduction
%%%%%%%%%%%%%%%%%%%%%%%%%
% general introduction & state-of-art
%\textcolor{red}{Introduction of DM and its direct detection - }
The existence of dark matter (DM) is supported by various astronomical and cosmological observations~\cite{clowe2004weak, rubin1980rotational, aghanim2018planck} but its nature remains unknown.
The most promising DM candidate is the so-called weakly interacting massive particle (WIMP)~\cite{wimp_theory1}, which explains the current abundance of dark matter as a thermal relic of the Big Bang~\cite{thermal_relic}. 
%and has mass larger than Lee-Weinberg limit~\cite{lee-weinberg_limit} of about 2\,GeV/c$^2$.
In the last three decades, numerous terrestrial experiments have been built to detect the faint interactions between WIMPs and ordinary matter. Among them, experiments using dual-phase (liquid/gas) xenon time projection chambers (TPCs) ~\cite{lux, pandax, xe1t_sr1} are leading the search for WIMPs with masses from a few GeV/c$^2$ to TeV/c$^2$. 
% Currently, the XENON1T experiment~\cite{xe1t_instrument} has set the most stringent limit on the WIMP-nucleon spin-independent elastic cross-section ($\sigma_{\mathrm{SI}}$) above 6\,GeV/c$^2$~\cite{xe1t_sr0, xe1t_sr1}.
%The XENON1T experiment~\cite{xe1t_instrument}, using a TPC filled with 3.2\,tonne of ultra-pure liquid xenon (LXe), is the first tonne-scale dark matter search experiment. 
%It has recently published the dark matter search results using data with one-tonne-year exposure~\cite{xe1t_sr1}, and reported the lowest background ever achieved in a dark matter search experiment~\cite{xe1t_analysis2}.
%The results set the most stringent limits on the WIMP-nucleon spin-independent elastic cross-section ($\sigma_{\mathrm{SI}}$) above 6 GeV/c$^2$.
The mass of the WIMP is expected to be larger than about 2\,GeV/c$^2$ from the Lee-Weinberg limit~\cite{thermal_relic} assuming a weak scale interaction.
On the other hand, DM in the sub-GeV/c$^2$ mass range has been proposed in several models~\cite{low_mass_dm, boehm2004can, boehm2004scalar}.
%On the other hand, hidden-sector DM~\cite{alexander2016dark, battaglieri2017us} can have new force carrier which allows for DM mass go below the Lee-Weinberg limit, which is also of great interest in the search for dark matter.
In this letter, we report on a probe of light DM-nucleon elastic interactions by looking for electronic recoils (ERs) in XENON1T, induced by secondary radiation (Bremsstrahlung~\cite{bremsstrahlung} and the Migdal effect~\cite{original_migdal, migdal}) that can accompany a nuclear recoil (NR).
ER signals induced by the Migdal effect and Bremsstrahlung (BREM) can go well below 1 keV, where the detection efficiency for scintillation signal is low. 
% Therefore, we add an approach that utilizes the ionization signal only (S2-only analysis), as well as both scintillation and ionization signals (S1-S2 analysis), which enables to lower the detection threshold. 
Therefore, in addition to the analysis utilizing both ionization and scintillation signals, we performed analysis using the ionization signal only, which improves the detection efficiency for sub-keV ER events.
We present results from a proble of light DM (LDM) with masses as low as 85 MeV/c$^2$.

% why light WIMP
% give introduction to Migdal effect and BREM
%\textcolor{red}{Intro to Xe1T and limitation on NR search - }
%The XENON1T direct dark matter detection experiment~\cite{xe1t_instrument} uses a dual-phase (liquid/gas) TPC containing 3.2\,tonne of ultra-pure liquid xenon (LXe), with 2\,tonne employed as the target material.
The XENON1T direct dark matter detection experiment~\cite{xe1t_instrument} uses a dual-phase TPC containing 2\,tonnes of ultra-pure liquid xenon (LXe) as the active target material.
It is located at the INFN Laboratori Nazionali del Gran Sasso (LNGS) in Italy, which has an average rock overburden of 3600\,m water-equivalent.
%Both S1 and S2 signals are detected by top and bottom arrays of 248 Hamamatsu R11410-21 3$^{\prime\prime}$ photomultiplier tubes (PMTs).
The prompt primary scintillation (S1) and secondary electroluminescence of ionized electrons (S2) signals are detected by top and bottom arrays of 248 Hamamatsu R11410-21 3$^{\prime\prime}$ photomultiplier tubes (PMTs)~\cite{pmt1, pmt2}.
They are used to reconstruct the deposited energy and the event interaction position in three dimensions, which allows for fiducialization of the active volume~\cite{xe1t_analysis1, xe1t_analysis2}.
% This enables for mitigating background events near the detector walls and also neutrons and gamma-rays that scatter multiple times within the active volume.
%This helps mitigating the backgrounds near the detector walls and neutrons which can scatter multiple times within the active volume.
The XENON1T experiment has published WIMP search results by looking for NRs from WIMP-nucleus elastic scattering using data from a one-tonne-year exposure, achieving the lowest ER background in a DM search experiment~\cite{xe1t_sr1}.
The excellent sensitivity of LXe experiments to heavy WIMPs comes from the heavy xenon nucleus which gives a coherent enhancement of the interaction cross-section and from the large NR energy.
%the allow for a large momentum transfer in WIMP-nucleus elastic interactions, resulting in a large recoil energy of the nucleus.
The sensitivity to sub-GeV/c$^2$ LDM, on the other hand, decreases rapidly with lowering DM mass since detectable scintillation and ionization signals produced by these NRs become too small.
The energy threshold (defined here as the energy at which the efficiency is 10\%) in a LXe TPC is mainly limited by the amount of detectable S1 signals.
A significant fraction of deposited NR energy is transferred into heat due to the Lindhard quenching effect~\cite{lindhard1963integral}.
Thus the detection efficiency for these NRs becomes extremely low, with less than 10\% for NRs below 3.5 keV in XENON1T~\cite{xe1t_sr1}.
It is challenging to detect the NR signals from LDM interactions.

%\textcolor{red}{Introduction of Migdal and BREM radiations - }
%In this letter, we report on a search for LDM-nucleon elastic interactions through looking for electronic recoils (ERs) induced by secondary radiations that can accompany an NR (BREM~\cite{BREM} and Migdal effect~\cite{original_migdal, migdal}).
Unlike NRs, ERs lose negligible energy as heat because recoil electrons have small masses compared with xenon nuclei.
This leads to a lower energy threshold for ER signals.
%, down to about 1\,keV in XENON1T~\cite{xe1t_sr1, xe1t_analysis2}.
%Although there is no discrimination against the major background from $\beta$ decay of $^{214}$Pb at the same energy, probing the ER signals induced by BREM and the Migdal effect enables a significant boost of the sensitivity to LDMs owing to the lowered threshold.  %in LXe detector.
Probing the ER signals induced by the Migdal effect and BREM enables a significant boost of XENON1T's sensitivity to LDMs, thanks to the lowered threshold.  %in LXe detector.

%Explanation of Migdal and BREM
%\textcolor{red}{Mechanism of Migdal and BREM radiations - }
When a particle elastically scatters off a xenon nucleus, the nucleus undergoes a sudden momentum change with respect to the orbital atomic electrons, resulting in the polarization of the recoiling atom and a kinematic boost of the electrons.
The de-polarization process can lead to BREM emission~\cite{bremsstrahlung}, and the kinematic boost of atomic electrons can result in  ionization and/or excitation of the atom, which eventually causes secondary radiation, known as the Migdal effect (MIGD)~\cite{original_migdal, migdal}.
%The differential rate of ER, $dR/dE_{\mathrm{e}}$, induced by Migdal radiations can be written as

The differential rate of BREM emission with photon energy E$_{\mathrm{ER}}$ is given by

%\begin{equation}
%    \frac{d^2 R}{d\omega dv} \propto \frac{\left | f(\omega) %\right|^2}{\omega}
%    \sqrt{1 - \frac{2\omega}{\mu_N v^2}}
%    \left(
%    1 - \frac{\omega}{\mu_N v^2}
%    \right),
%    \label{eq:brem}
%\end{equation}
\begin{equation}
    \frac{d^2 R}{dE_{\mathrm{ER}} dv} \propto \frac{\left | f(E_{\mathrm{ER}}) \right|^2}{E_{\mathrm{ER}}}
    \sqrt{1 - \frac{2E_{\mathrm{ER}}}{\mu_N v^2}}
    \left(
    1 - \frac{E_{\mathrm{ER}}}{\mu_N v^2}
    \right),
    \label{eq:brem}
\end{equation}
where $v$, $\mu_N$, and $f(E_{\mathrm{ER}})$ are the velocity of DM, the reduced mass of the xenon nucleus and DM, and the atomic scattering factor, respectively~\cite{bremsstrahlung}.

\begin{figure}[htpb]
    \centering
    \includegraphics[width=0.9\columnwidth]{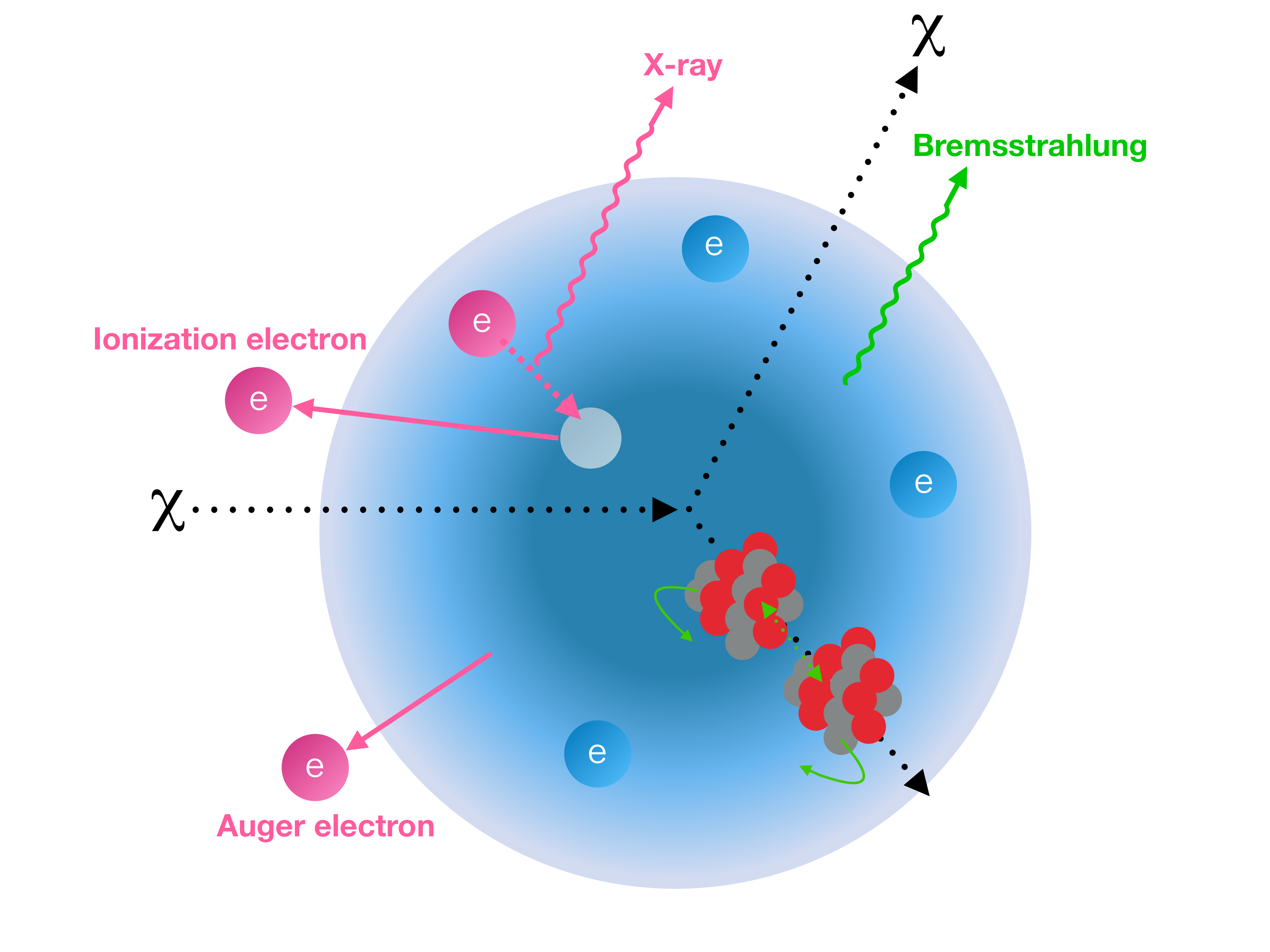}
    \caption{Illustration of the ER signal production from BREM (green) and MIGD processes (pink) after elastic scattering between DM ($\chi$) and a xenon nucleus. 
    The electrons illustrated in pink represent those involved in ionization, de-excitation, and Auger electron emission during a MIGD process.
    % In both cases, additional electronic recoil signals are produced, which significantly extend the sensitivity to sub-GeV WIMPs.
    }
    \label{fig:migdal_diagram}
\end{figure}

The differential rate of MIGD process giving an NR of energy E$_{\mathrm{NR}}$ accompanied by an ER of energy E$_{\mathrm{ER}}$ is given by

%\begin{widetext}
\begin{equation}\large
%    \frac{dR}{dE_{\mathrm{ER}}} \simeq \int dE_R d\nu_{\mathrm{DM}} \frac{dR}{dE_R d\nu_{\mathrm{DM}}} \left( \\
%    \begin{array}{cl}
%        1/(2\pi) & \sum_{n,l} \frac{d}{d (E_{\mathrm{ER}}-E_{nl})}} p^c_{q_e} (n,l\rightarrow (E_{\mathrm{ER}} - E_{nl}) )\\
%        + & \sum_{n,l,n^{\prime},l^{\prime}} p^d_{q_e} (n,l\rightarrow n^{\prime}, l^{\prime}) \delta (E_{\mathrm{ER}}-E_{n,l}+ E_{n^{\prime}, l^{\prime}}) \\
%    \end{array}
%    \right)
\begin{array}{lll}
    \frac{d R}{dE_{\mathrm{ER}}} &\simeq & \int dE_{\mathrm{NR}} dv \frac{d^{2}R}{dE_{\mathrm{NR}}dv}  \\
     && \times \frac{1}{2\pi} \sum_{n,l} \frac{d}{dE_{\mathrm{ER}}} p^c_{q_e} (n,l\rightarrow E_{\mathrm{ER}}-E_{n,l}),
\end{array}
\label{eqn:migdal_effect}
\end{equation}
%\end{widetext}
where $p^c_{q_e}$ is the probability for an atomic electron, with quantum numbers ($n$, $l$) and binding energy $E_{n,l}$, to be ionized and receive a kinetic energy $E_{\mathrm{ER}}-E_{n,l}$~\cite{migdal}.
$p^c_{q_e}$ is related to $q_{e}$ which is the momentum of each electron in the rest frame of the nucleus after the scattering.
%The shell vacancy is immediately refilled and X-ray with energy of $E_{n,l}$ is emitted, which is not distinguished with the energy deposited by the ionized electron.
The shell vacancy is immediately refilled, and an X-ray or an Auger electron with energy $E_{n,l}$ is emitted.
$E_{n,l}$ is measured simultaneously with the energy deposited by the ionized electron, since the typical timescale of the de-excitation process is  $\mathcal{O}$(10) fs.
Atomic electrons can also undergo excitation instead of ionization, in which case an X-ray is emitted during de-excitation~\cite{migdal}.
Excitation, however, is sub-dominant compared to the ionization process, and thus is not considered in this analysis.
%The first and second terms in the brackets represent the differential rates of ionizing a shell electron with quantum number of ($n$, $l$), which is followed by a characteristic X-ray for shell vacancy refilling, and exciting ($n$, $l$) to ($n^{\prime}$, $l^{\prime}$) quantum state, respectively.
%$E_{n,l}$ and $E_{n^{\prime},l^{\prime}}$ are the binding energies of the shell electrons.
%$p^c_{q_e}$ and $p^d_{q_e}$ are the transition probabilities for the ionization and excitation, respectively, and are given in recent results from~\cite{migdal} which performs a detailed numerical calculation of the Migdal effect in LXe.
%The calculation of the signal rate in~\cite{migdal} is based on the isolated atom assumption, which has large uncertainty in the calculation of valence electrons ($n$=5) in LXe. 
%Thus, signals from the Migdal effect, where the valence electrons are ionized, are neglected in this work. 
Only the contributions from the ionization of M-shell ($n$=3) and N-shell ($n$=4) electrons are considered in this work, as inner electrons ($n$$\leq$2) are too strongly bound to the nucleus to contribute significantly.
The contribution from the ionization of valence electrons ($n$=5) is neglected because it is subdominant in region of interest compared to the ones from M- and N-shell electrons, and the calculation of it has large uncertainty since the assumption of isolated atom is used for LXe~\cite{migdal}.
An illustration of MIGD and BREM is given in Fig.~\ref{fig:migdal_diagram}.
The radiation from MIGD is typically 3-4 orders of magnitude more likely to occur than BREM.
%The signal rate from the Migdal effect is 3-5 orders of magnitude smaller than that of NR for elastic DM-nucleus scattering.
Although only a very small fraction (about 3$\times 10^{-8}$ and 8$\times 10^{-6}$ for DM masses of 0.1 and 1.0 GeV/c$^2$, respectively) of NRs accompanies MIGD radiations, the larger energy and ER nature make them easier to be detected than the pure NRs.
% Therefore, we only report on a search for sub-GeV WIMPs by probing the electronic recoils induced by the Migdal effect in this work.
% However, it is worth noting that despite the well-motivated mechanism of the Migdal effect, such signature has not yet been experimentally confirmed.

The data used in previous analyses~\cite{xe1t_sr1} consists of two science runs with a livetime of 32.1 days (SR0) and 246.7 days (SR1), respectively.
The two runs were taken under slightly different detector conditions.
% The drift field was decreased from 120\,V/cm in SR0 to 81\,V/cm in SR1, after an interruption by an earthquake.
% The impurity concentration in LXe and the self-trigger threshold of the digitizers~\cite{xe1t_daq} for each PMT channel were slightly higher for a portion of SR0 compared with SR1, both of which result in a lower detection efficiency~\cite{xe1t_analysis1}.
% Compared with the search for NR signals in XENON1T~\cite{xe1t_sr1}, a search for ER signals from the Migdal effect or BREM is more strongly affected by the dominant ER background from $\beta$ decays of $^{214}$Pb~\cite{xe1t_analysis2}, which are uniformly distributed in the sensitive target.
% Therefore, taking SR0 data into account would not result in a significant sensitivity increase.
% % Therefore, the sensitivity doesn't increase significantly with SR0 data taken into account.
% We thus use SR1 data only in this work.
To maximize the amount of data acquired under stable detector conditions we decided to use SR1 only.
%SR0 data are removed from this analysis because of its negligible contribution to sensitivity.
% Therefore adding SR0 data does not have a significant improvement in sensitivity and, thus, are not used in this work.
The same event selection, fiducial mass, correction, and background models as described in~\cite{xe1t_sr1} are used for the SR1 data, which we refer to as the S1-S2 data in later text.
The exposure of the S1-S2 data is about 320\,tonne-days.
The interpretation of such S1-S2 analysis is based on the corrected S1 (cS1) signal and the corrected S2 signal from the PMTs at the bottom of the TPC (cS2$_b$).
%, to minimize the signal variation due to geometric light-collection effects~\cite{xe1t_analysis1}.

\begin{figure}[htbp]
\centering
\includegraphics[width=\columnwidth]{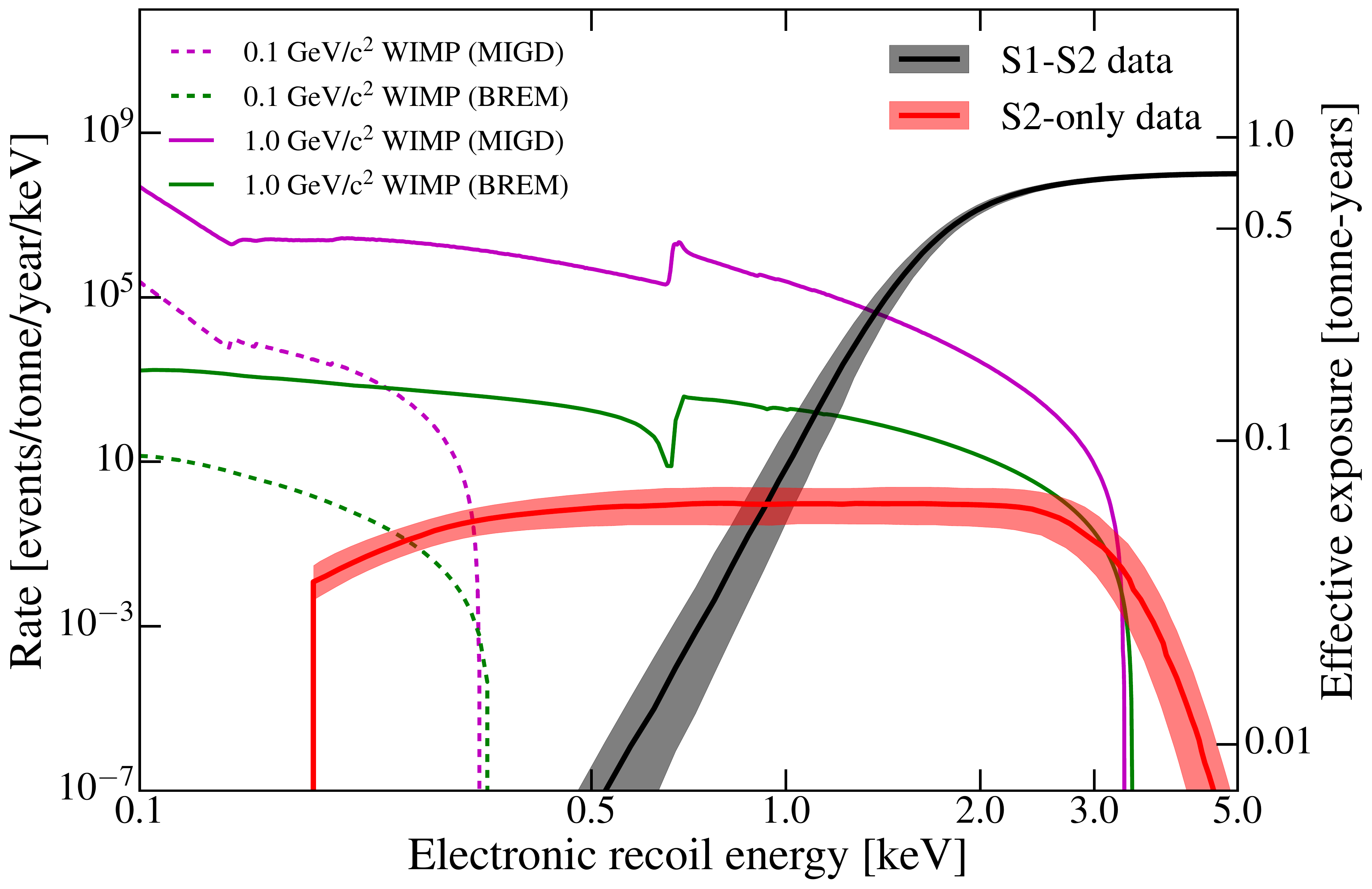}
\caption{Median effective exposures of ER signals after event selections as a function of recoil energy for the S1-S2 data (black line) and S2-only data (red line).
% The efficiency of S2-only data includes the effect of the event selection on the interaction position in order to maximize the signal-to-background ratio, different from the efficiency of the S1-S2 data which is estimated in fixed 1.3-tonne fiducial volume. 
% The details are in~\cite{xe1t_s2only}.
The 68\% credible regions of the effective exposures are also shown as the shaded regions.
% As a reference, the 68\% credible regions of detection efficiencies in the top and bottom 20\,cm detector volumes are shown in blue and red shaded regions, respectively.
The expected event rate of DM-nucleus scattering from MIGD/BREM for DM masses of 0.1 and 1.0\,GeV/c$^2$ are overlaid as well, in magenta/green dashed and solid lines, respectively, assuming a spin-independent DM-nucleon interaction cross section of 10$^{-35}$\,cm$^2$.
}
\label{fig:efficiency}
\end{figure}

\begin{figure}[htpb]
    \centering
    \includegraphics[width=\columnwidth]{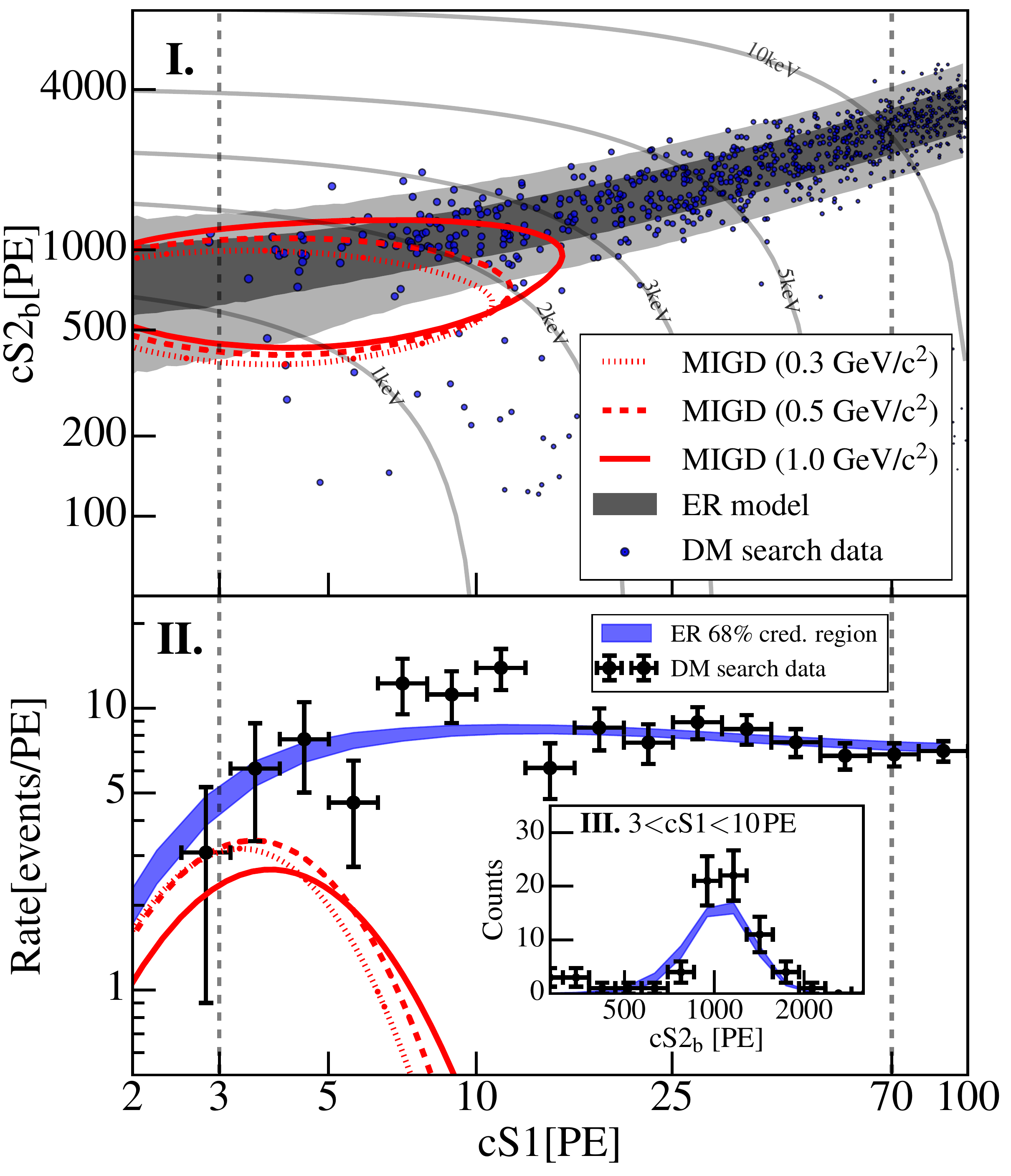}
    \caption{Comparison of cS1 and cS2$_b$ spectra between the S1-S2 data and the signal response model~\cite{xe1t_analysis2}.
    In the upper panel (I), the distribution of the S1-S2 data in (cS2$_b$, cS1) space is shown as light blue dots, along with the best-fit ER background model (black shaded region).
    The contours containing 90\% of the expected signals from MIGD for 0.3, 0.5, and 1\,GeV/c$^2$ DM are shown in red dotted, dashed, and solid lines, respectively.
    Gray lines show isoenergy contours in ER energy.
    The events having lower cS2$_b$ than what we expect for ER are mostly surface backgrounds~\cite{xe1t_sr1}, which have minimal impact to the results of this study.
    The lower panel (II) shows the projected cS1 distribution of the S1-S2 data, where cS2$_b$ is within the 2$\sigma$ contour of ER model shown in panel (I). For comparision, the 68\% credible region of cS1 distribution from ER background model (blue shadow) is shown, which is mainly attributed to the systematic uncertainties of the model.
    The cS1 distributions of the expected signals from MIGD for 0.3, 0.5, and 1\,GeV/c$^2$ DM with assumed spin-independent DM-nucleon cross sections of 2$\times$10$^{-28}$, 10$^{-36}$, and 10$^{-38}$ cm$^2$, respectively, are shown as well.
    The vertical dashed lines indicate the region of interest (3-70\,PE). 
    The inset, panel (III) shows the cS2$_b$ distribution, with cS1 in (3, 10)\,PE, compared with the 68\% credible region of the cS2$_b$ spectra from the ER background model (blue shadow).
    }
    \label{fig:rn220_fit}
\end{figure}

%\textcolor{red}{Intro to standard (cS2b, cS1) data - }
The region of interest in the S1-S2 data is from 3 to 70 photoelectrons (PEs) in cS1, which corresponds to median ER energies from 1.4 to 10.6\,keV in the 1.3-tonne fiducial volume (FV) of XENON1T.
The lower value is dictated by the requirement of the 3-fold PMT coincidence for defining a valid S1 signal~\cite{xe1t_analysis1}. %which is the dominant limitation on the detection efficiency of the search using the S1-S2 data.
A detailed signal response model~\cite{xe1t_analysis2} is used to derive the influence of various detector features, including the requirement of the 3-fold PMT coincidence, on the reconstructed signals.
The effective exposure, which is defined as exposure times detection efficiency, and its uncertainty as a function of deposited ER energy for the S1-S2 data are shown in Fig.~\ref{fig:efficiency}, with the signal spectra from MIGD and BREM induced by 0.1\,GeV/c$^2$ and 1\,GeV/c$^2$ DM masses overlaid.
The (cS2$_b$, cS1) distribution of S1-S2 data are shown in Fig.~\ref{fig:rn220_fit}.
The rise of the event rate at around 0.85\,keV for DM mass of 1.0\,GeV/c$^2$ is contributed by the ionization of M-shell electrons~\cite{migdal, bremsstrahlung}. 
% The ER spectra from the  and Migdal effect induced by 0.1\,GeV/c$^2$ and 1\,GeV/c$^2$ WIMPs are overlaid in Fig.~\ref{fig:efficiency} as well.
% Due to the non-linear signal response in the very low-energy region, the detection efficiency is a function of deposited ER energy, which is shown in Fig.~\ref{fig:efficiency}.
% The light collection efficiency in the XENON1T detector has a large spatial dependence. This dependence is measured with $^{83\mathrm{m}}$Kr calibration data~\cite{xe1t_analysis1}, and is taken into account in the model~\cite{xe1t_analysis2}.
% We show the 68\% credible regions of the detection efficiencies in the top and bottom 20\,cm regions of the detector in Fig.~\ref{fig:efficiency} as a reference.
In our signal models, deposited energy below 1\,keV, at which the median detection efficiency in 1.3-tonne FV is 10\%, from MIGD and BREM is neglected for the S1-S2 data in the following analysis.
There are only two sub-keV measurement of ionization yield for ER in LXe~\cite{xe127_lux, boulton2017calibration}.% and our signal response model~\cite{xe1t_analysis2} is more conservative than their measurements. %since we do not have dedicated calibration sources in the energy region below 1\,keV.

The S1-S2 data selections~\cite{xe1t_analysis1} provide excellent rejection of noise and backgrounds, and are characterized as well by the well-established background models~\cite{xe1t_analysis2} and a fully blind analysis~\cite{xe1t_sr1}.
However they also limit the detection efficiency of $\mathcal{O}$(1) keV energy depositions.
We therefore consider also the events with no specific requirement on S1 (S2-only data) in this work.
Although the reduction of available information in the S2-only data implies less background discrimination, the increased detection efficiency in the $<$ 1\,keV ER energy region, shown in Fig.~\ref{fig:efficiency}, enables a more sensitive search for LDM-nucleus interactions through MIGD and BREM.
The interpretation of such S2-only data is based on the uncorrected S2 signal, combining both signals from top and bottom PMT arrays.
%since the depth of interaction ($z$ position) cannot be reconstructed without S1.

\begin{figure}
    \centering
    \includegraphics[width=\columnwidth]{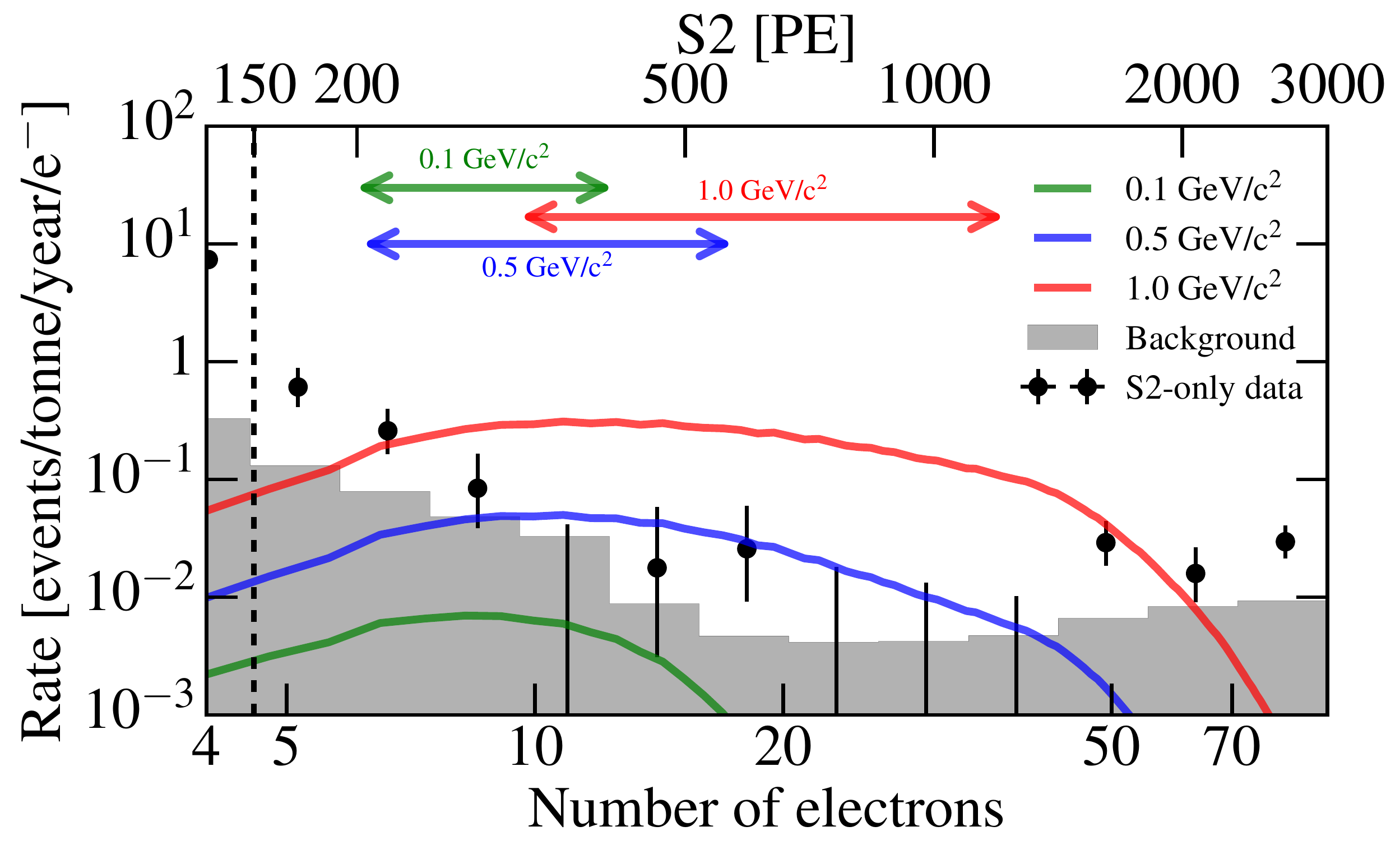}
    \caption{
    % \textcolor{red}{[Data and background from plot grabbing]}
    Observed S2 spectra for the S2-only data after the optimized selection described in~\cite{xe1t_s2only}.
    The expected spectra of ER signals induced by MIGD for DM with mass of 0.1, 0.5, and 1.0\,GeV/c$^2$ are shown in green, blue, and red solid lines, respectively, assuming the spin-independent DM-nucleon interaction cross section of 1.2$\times$10$^{-37}$, 1.5$\times$10$^{-39}$, and  2.0$\times$10$^{-39}$\,cm$^2$ for 0.1, 0.5, and 1.0\,GeV/c$^2$ DM, respectively.
    The gray shaded region shows the conservative background model used in analysis of S2-only data.
    The arrows indicate the S2 ROIs that are later used in inference for the three DM signals above-mentioned.
    The S2 threshold used for the S2-only data is denoted in the dashed black line.
    }
    \label{fig:s2onlydata}
\end{figure}

%\textcolor{red}{Optimal S2-only selection}
We analyze the S2-only data as in \cite{xe1t_s2only}, using the LDM signal models appropriate for MIGD and BREM. As detailed in \cite{xe1t_s2only}, 30\% of the data was used for choosing regions of interest (ROIs) in S2 and event selections. A different S2 ROI is chosen for each dark matter model and mass to maximize the signal-to-noise ratio, based on the training data. The event selections used for this work are the same as in \cite{xe1t_s2only}, and mainly based on the width of each S2 waveform, reconstructed radius, and PMT hit-pattern of the S2.
Fig.~\ref{fig:s2onlydata} shows the observed S2 spectra for the S2-only data, along with the expected DM signal distributions by MIGD with masses of 0.1, 0.5, and 1.0\,GeV/c$^2$, respectively.
The S2 ROIs for these three DM models shown in Fig.~\ref{fig:s2onlydata} are indicated by the colored arrows.
Conservative estimates of the background from $^{214}$Pb-induced $\beta$ decays, solar-neutrino induced NRs, and surface backgrounds from the cathode electrode are used in the inference~\cite{xe1t_s2only}.
The background model is shown in Fig.~\ref{fig:s2onlydata} as shaded gray region.

% Signal model of standard data and S2-only data
% and the light/charge yield discussion
%\textcolor{red}{Signal model - }
The detector response to ERs from MIGD and BREM in (cS2$_b$, cS1) space (for the S1-S2 data) and in reconstructed number of electrons (for the S2-only data) is derived using the signal response model described in~\cite{xe1t_analysis2}.
% The signal response model, which integrates our best knowledge of the scintillation and ionization yield of LXe as well as the reconstruction efficiency of the signals, is constrained by fitting to the $^{220}$Rn calibration data~\cite{rn220_xe100}.
% Excellent agreement between the model expectation and $^{220}$Rn calibration data is observed~\cite{xe1t_analysis2}.
% The ionization yield below 1\,keV is estimated based on~\cite{xe1t_analysis2} and used in interpreting the S2-only data.
%Note that the ionization yield used in this work is more conservative than the Noble Element Simulation Technique (NEST) v2 model~\cite{nest_v2}, which uses the only measurements~\cite{xe127_lux} of ionization yield for ERs so far in LXe below 1\,keV.
Note that the ionization yield used for the S2-only data is more conservative than the Noble Element Simulation Technique (NEST) v2 model~\cite{nest_v2}.
Fig.~\ref{fig:rn220_fit} shows the comparison between the expectation from our signal response model and the S1-S2 data, as well as the (cS2$_b$, cS1) distribution of ERs from MIGD. 
Signal contours for different DM masses are similar since the energy spectra from MIGD and BREM are not sensitive to incident dark matter velocity as long as it is kinematically allowed.
%Fig.~\ref{fig:s2onlydata} shows the S2-only events after event selections in the traditional approach, along with the expected spectra of ERs induced by the Migdal effect from 0.1, 0.3, and 0.5\,GeV/c$^2$ DM masses.
% A threshold of 15 electrons is applied to S2-only data, which corresponds to approximately the electron number of lowest deposited ER energy measured~\cite{xe127_lux}.
%The responses of ERs induced by  are similar to those induced by the Migdal effect, and are not shown.
We have ignored the contribution of NRs in the signal model of MIGD and BREM, since it is small compared with ERs from MIGD and BREM in this analysis and there is no measurement of scintillation and ionization yields in LXe for simultaneous ER and NR energy depositions.
% This treatment gives more a conservative upper-limit for inferring the S2-only data.
We use the inference only for DM mass below 2\,GeV/c$^2$, above which the contribution of an NR in the signal rate becomes comparable with or exceeds the signal model uncertainty.

\begin{figure}[tpb]
    \centering
    \includegraphics[width=\columnwidth]{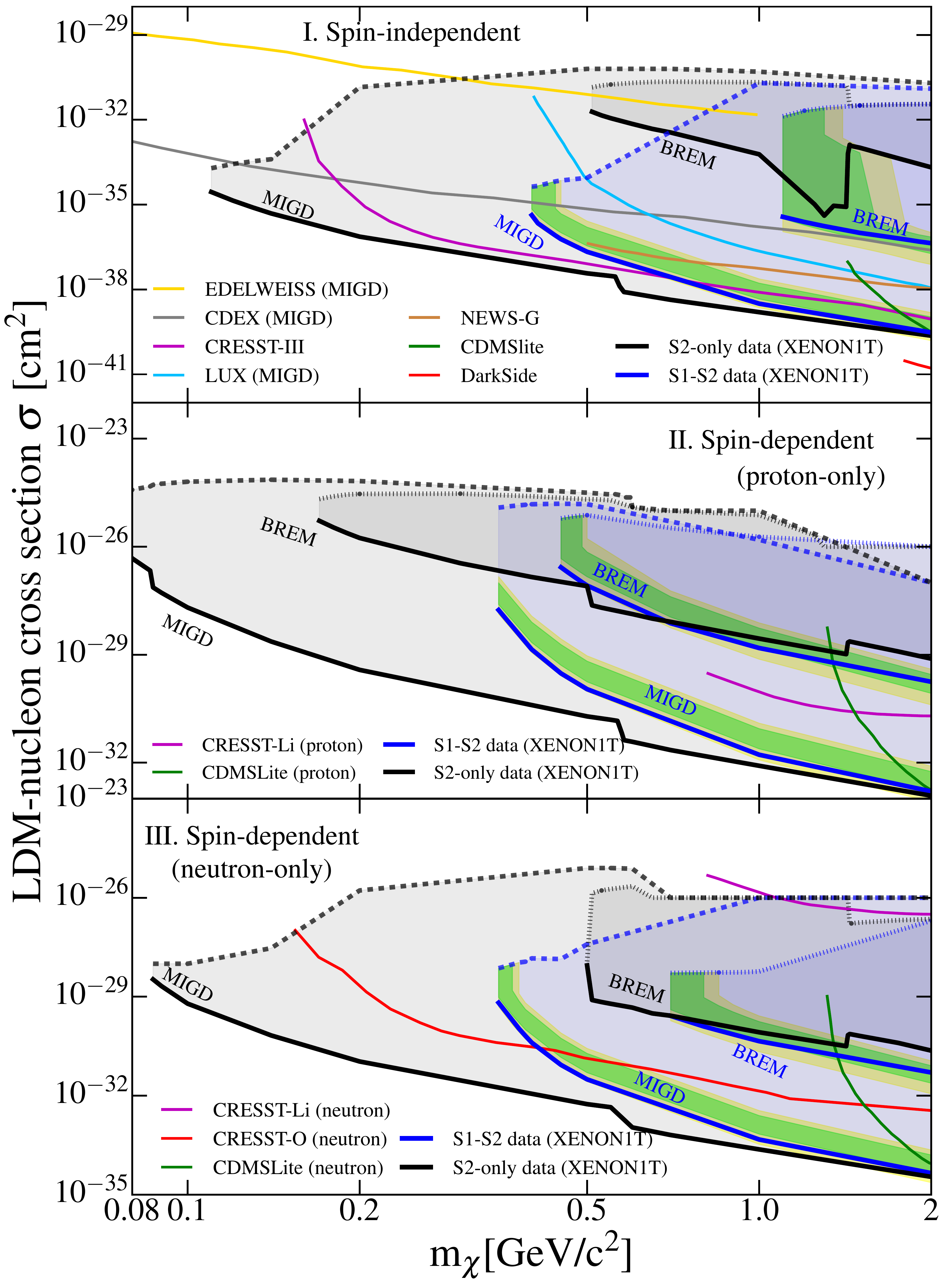}
    \caption{
    Limits on the SI (upper panel), SD proton-only (middle panel), and SD neutron-only (lower panel) DM-nucleon interaction cross-sections at 90\% C.L. using signal models from MIGD and BREM in the XENON1T experiment with the S1-S2 data (blue contours and lines) and S2-only data (black contours and lines). 
    The solid and dashed (dotted) lines represent the lower boundaries (also referred to as upper limits) and MIGD (BREM) upper boundaries of the excluded parameter regions.
    Green and yellow shaded regions give the 1 and 2\,$\sigma$ sensitivity contours for upper limits derived using the S1-S2 data, respectively.
    The upper limits on the SI DM-nucleon interaction cross sections from LUX~\cite{lux_migdal},  EDELWEISS~\cite{edelweiss}, CDEX~\cite{cdex}, CRESST-III~\cite{cresst_iii}, NEWS-G~\cite{news_g}, CDMSLite-II~\cite{cdmslite_ii}, and DarkSide-50~\cite{darkside_s2only}, and upper limits on the SD DM-nucleon interaction cross sections from CRESST~\cite{cresst_iii, cresst_sd} and CDMSLite~\cite{cdmslite_sd} are also shown.
    Note that the limits derived using the S1-S2 and S2-only data are inferred using unbinned profile likelihood method~\cite{xe1t_analysis1} and simple Poisson statistics with the optimized event selection~\cite{xe1t_s2only}, respectively.
    % As in~\cite{xe1t_s2only}, the jumps in the S2-only limits are due to changes in the observed event count due to the mass-dependent ROIs.
    The sensitivity contours for the S2-only data is not given since the background models used in the S2-only data are conservative~\cite{xe1t_s2only}.
    }
    \label{fig:migdal_inference}
\end{figure}

%\textcolor{red}{Inference and Results - }
The S1-S2 data are interpreted using an unbinned profile likelihood ratio as the test statistic, as detailed in~\cite{xe1t_analysis2}.
The unbinned profile likelihood is calculated using background models defined in cS2$_b$, cS1, and spatial coordinates.
The uncertainties from the scintillation and ionization yields of ER backgrounds, along with the uncertainties in the estimated rates of each background component, are taken into account in the inference~\cite{xe1t_analysis2}.
% Since we don't have a solid background model for the S2-only data, all the observed data is considered as signal events and a conservative exclusion limit is calculated. In this work, the optimal interval method~\cite{optimal_interval}, that can maximize the exclusion sensitivity in the case of unknown background, is used to calculate an exclusion limit for S2-only data (conservative approach).
%The optimized approach for inferring the S2-only data, with the optimal data selection and background subtraction, is detailed in~\cite{xe1t_s2only}.
%In the traditional S2-only approach, we consider all the observed data in the region of interest as signal. 
%The optimal interval method~\cite{optimal_interval}, which gives an optimized upper limit in the presence of unknown background, is used for inferring the S2-only data.
The inference procedure for the S2-only data is detailed in~\cite{xe1t_s2only}, which is based on simple Poisson statistics using the number of events in the S2 ROI.
%For the traditional S2-only approach, we consider all observed data in the region of interest as signal, and use the optimal interval method~\cite{optimal_interval} to give an optimized upper limit in the presence of unknown background.
%The event rate of WIMP-nucleon elastic scattering is calculated assuming a standard isothermal WIMP halo with the asymptotic velocity of the local system of $v_{0}$ = 220 km/s, a local WIMP density of $\rho_{0}$ = 0.3 GeV/cm$^{3}$, the galactic escape velocity $v_{esc}$ = 544 km/s, and the Helm form factor for the nuclear cross section~\cite{astro_dm}.
The event rates of spin-independent (SI) and -dependent (SD) DM-nucleon elastic scattering are calculated following the approaches described in~\cite{astro_dm, xe1t_sr1} and~\cite{xe1t_sd}, respectively.

The results are also interpreted in a scenario where LDM interacts with the nucleon through a scalar force mediator $\phi$ with equal effective couplings to the proton and neutron as in the SI DM-nucleon elastic scattering.
In this scenario, the differential event rates are corrected by ${m_\phi}^4 / ({m_\phi}^2 + q^2 / c^2)^2$ ~\cite{lightmediator, pandax_lm}, where $q = \sqrt{2 m_N E_R}$ and $m_N$ are the momentum transfer and the nuclear mass, respectively.
We take the light mediator (LM) regime where the momentum transfer is much larger than ${m_\phi}$ and thus the interaction cross section scales with $m_\phi^4$.
In this regime, the contribution of NRs is largely suppressed compared with SI DM-nucleon elastic scattering due to the long-range nature of the interaction. Therefore, the results are interpreted for DM mass up to 5 GeV/c$^2$ for SI-LM DM-nucleon elastic scattering.
% We scanned the LDM masses ranging from 0.075\,GeV to 2\,GeV for the search of ER signals induced by the Migdal effect.
%No significant sub-GeV/c$^2$ DM signal is indicated in the search using the standard data.

%%%%%%%%%%%%%%%%%%%%
%% Earth shielding effect
%%%%%%%%%%%%%%%%%%%%
% \textcolor{red}{Earth shielding effect - }
In addition, we also take into account the fact that DM particle may be stopped or scatter multiple times when passing through Earth's atmosphere, mantle, and core before reaching the detector (Earth-shielding effect)~\cite{earth_shielding_ref1, earth_shielding_ref2, earth_shielding_ref3}.
If the DM-matter interaction is sufficiently strong, the sensitivity for detecting such DM particles in terrestrial detectors, especially in underground laboratory, can be reduced or even lost totally.
Following~\cite{edelweiss}, \textit{verne} code~\cite{verne} is used to calculate the Earth-shielding effect for SI DM-nucleon interaction.
A modification of the \textit{verne} code based on the methodology in~\cite{verne_method} is applied for the calculations of SD and SD-LM DM-nucleon interactions.
To account for the Earth-shielding effect for SD DM-nucleon interaction, $^{14}$N in the atmosphere and $^{29}$Si in Earth's mantle and core are considered, and their spin expectation values, $\langle S_n \rangle$ and $\langle S_p \rangle$, are taken from~\cite{sd_earth_shielding_ref}.
%We report only the lower boundaries of excluded parameter space (referred to as upper limit in later context) but not the upper boundaries since they are not of the interest in this work.
%In the results of this work, we show only the lower boundaries of excluded parameter space (referred to as upper limit in later context) but not the upper boundaries since they are not of the interest of this study. The upper boundaries can be found in supplementary materials.
Both the lower and upper boundaries of excluded parameter space are reported in this work.
The lower boundaries are conventionally referred to as upper limits in later context, and are the primary interest of this work.
The upper boundaries are dominated by the overburden configuration of the Gran Sasso laboratory which hosts the detector.

\begin{figure}[htpb]
 \centering
 %\begin{center}
    \includegraphics[width=\columnwidth]{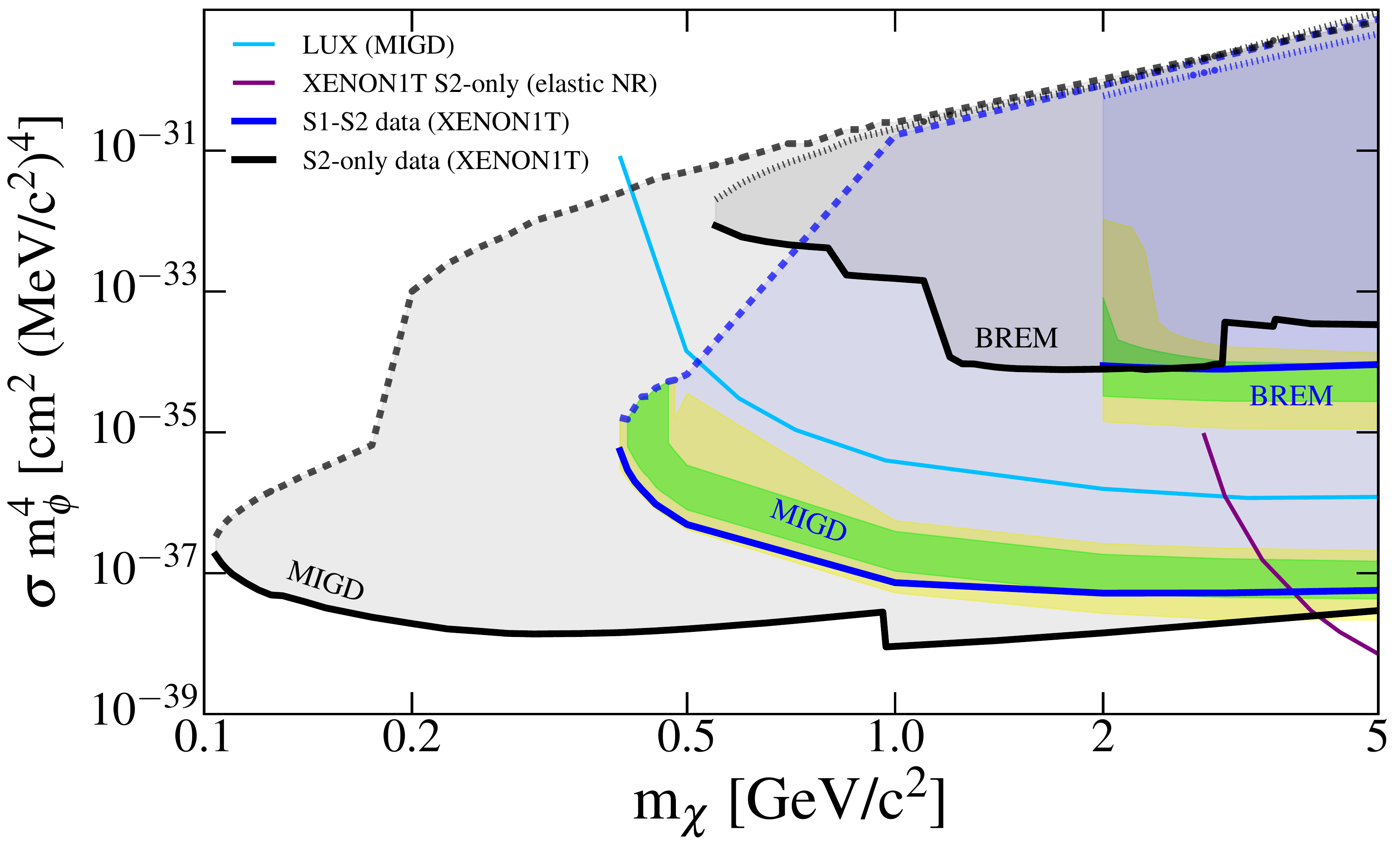}
    \caption{Limits on the SI-LM DM-nucleon interaction cross-sections at 90\% C.L. using signal models from MIGD and BREM in the XENON1T experiment with the S1-S2 data (blue contours and lines) and S2-only data (black contours and lines). 
    The figure description is the same as in Fig.~\ref{fig:migdal_inference}.
    The upper limits on the SI DM-nucleon interaction cross sections from LUX~\cite{lux_migdal} and XENON1T S2-only (elastic NR results)~\cite{xe1t_s2only} are also shown.
    }
    \label{fig:migdal_inference_lm}
 %\end{center}
\end{figure}

No significant excess is observed above the background expectation in the search using the S1-S2 data.
%The 90\% confidence-level (C.L.) upper limits on the SI and SD (proton-only and neutron-only cases) DM-nucleon interaction cross-section using signal models from the  and Migdal effect with masses from 0.075\,GeV/c$^2$ to 2\,GeV/c$^2$ at 90\% C.L. are given in Fig.~\ref{fig:migdal_inference}.
Fig.~\ref{fig:migdal_inference} shows the 90\% confidence-level (C.L.) limits on the SI and SD (proton-only and neutron-only cases) DM-nucleon interaction cross-section using signal models from MIGD and BREM with masses from about 85\,MeV/c$^2$ to 2\,GeV/c$^2$, and Fig.~\ref{fig:migdal_inference_lm} shows the 90\% C.L. limits on the SI-LM DM-nucleon interaction cross-section with masses from about 100\,MeV/c$^2$ to 5\,GeV/c$^2$.
% along with the upper limits from LUX~\cite{lux_migdal}, EDELWEISS~\cite{edelweiss}, CDEX~\cite{cdex}, CRESST-II~\cite{cresst_ii}, CRESST-III~\cite{cresst_iii}, CRESST-surface~\cite{cresst_surface}, NEWS-G~\cite{news_g},  CDMSLite-II~\cite{cdmslite_ii}, and DarkSide-50~\cite{darkside_s2only}. 
% For LDM masses ranging from 0.3\,GeV and 0.5\,GeV, XENON1T SR1 data gave the most stringent constrain to the spin-independent WIMP-nucleon interaction cross section.
% The best upper limits for WIMP mass of 0.1, 0.5, and 1\,GeV through searching for Migdal radiations are 3.6$\times$10$^{-36}$, 2.2$\times$10$^{-38}$, and 3.1$\times$10$^{-39}$cm$^2$, respectively.
The sensitivity contours for the results derived using S2-only data are not shown because of the conservativeness of the background model. 
The upper limits derived using the S1-S2 data deviate from the median sensitivity by about 1-2\,$\sigma$ due to the under-fluctuation of the ER background in the low energy region.
As described in~\cite{xe1t_s2only}, the jumps in the S2-only limits are originating from the changes in the observed number of events due to the mass-dependent S2 ROIs.
The results, by searching for ER signals induced by MIGD, give the best lower exclusion boundaries on SI, SD proton-only, SD neutron-only, and SI-LM DM-nucleon interaction cross-section for mass below about 1.8, 2.0, 2.0, and 4.0\,GeV/c$^2$, respectively as compared to previous experiments~\cite{lux_migdal, edelweiss, cdex, cresst_iii, news_g, cdmslite_ii, darkside_s2only, cresst_sd, cdmslite_sd}.
The upper limits derived from the S1-S2 data become comparable with those from the S2-only data at $\sim$\,GeV/c$^2$ since the efficiency of the S1-S2 data to DM signals with mass of $\sim$\,GeV/c$^2$ becomes sufficiently high.
However, the upper limits derived from the S1-S2 data do not provide significantly better constraints than those from the S2-only data for DM masses larger than 1\,GeV/c$^2$, because both data are dominated by the ER background, which is very similar to the expected DM signal.

% The conditional probabilities of observing data $D$ under null hypothesis $p(D|H_0)$ and signal hypothesis $p(D|H_1)$ in the Bayesian inference are based on the extended unbinned likelihood.
% The critical probability is defined as:
% \begin{equation}
%     \small
%     p(H_0|D) = \frac{\int p(D|H_0) p(\theta) d\theta}{\int p(D|H_0) p(\theta) d\theta + \int p(D|H_1) p(S) p(\theta) d\theta dS},
%     \label{equ:critical_prob}
% \end{equation}
% where $\theta$ is the nuisance parameters which are integrated over in the Bayesian inference.
% The $p(\theta)$ is the prior of the nuisance parameters used in $H_0$ and $H_1$, which are mostly detector signal response related and has the constrain from the model testing on calibration data.
% The rest of nuisance parameter constrain comes from the rates of different background components, which are summarized in Table~\ref{tab:background_prior}.
% Astronomical parameters, that are related to WIMP flux calculation, of mean earth velocity, standard deviation of Maxwellian velocity distribution, Milky Way escaping velocity are fixed to xx, xx, and xx, respectively, in the inference.
% $S$ is the WIMP signal rate and has a uniform prior $p(S)$ in this study.
%%%%%%%%%%%%%%%%%%%%%%%%%
% Discussion/Summary
%%%%%%%%%%%%%%%%%%%%%%%%%
%\textcolor{red}{Summary - }
%In summary, we performed a search for LDMs by probing ER signals induced by the  and Migdal effect, using data from the XENON1T experiment. 
%ER signals through the new detection channels significantly enhance the sensitivity to masses previously unreachable through the standard NR searches. 
In summary, we performed a search for LDM by probing ER signals induced by MIGD and BREM, using data from the XENON1T experiment. 
These new detection channels significantly enhance the sensitivity of LXe experiments to masses unreachable in the standard NR searches. 
We set the most stringent upper limits on the SI and SD DM-nucleon interaction cross-sections for masses below 1.8\,GeV/c$^2$ and 2\,GeV/c$^2$, respectively.
%Together with the standard NR search~\cite{xe1t_sr1}, XENON1T results have reached unprecedented sensitivities to both low-mass (sub-GeV/c$^2$) and high-mass ($>$5\,GeV/c$^2$) DMs.
Together with the standard NR search~\cite{xe1t_sr1}, XENON1T results have reached unprecedented sensitivities to both low-mass (sub-GeV/c$^2$) and high-mass (GeV/c$^2$ - TeV/c$^2$) DM.
With the upgrade to XENONnT, we expect to further improve the sensitivity to DM with masses ranging from about 85\,MeV/c$^2$ to beyond a TeV/c$^2$.
% \textcolor{red}{Need some extension to increase the significance.}

%%%%%%%%%%%%%%%%%%%%%%%%%
% Acknowledgement
%%%%%%%%%%%%%%%%%%%%%%%%%
The authors would like to thank Masahiro Ibe and Yutaro Shoji for helpful discussions on MIGD and for providing us with the code for calculating the rate of MIGD radiation in xenon.
We would like to thank Bradley Kavanagh for helpful discussion on the Earth-shielding effect.
We gratefully acknowledge support from the National Science Foundation, Swiss National Science Foundation, German Ministry for Education and Research, Max Planck Gesellschaft, Deutsche Forschungsgemeinschaft, Netherlands Organisation for Scientific Research (NWO), Netherlands eScience Center (NLeSC) with the support of the SURF Cooperative, Weizmann Institute of Science, Israeli Centers Of Research Excellence (I-CORE), Pazy-Vatat, Fundacao para a Ciencia e a Tecnologia, Region des Pays de la Loire, Knut and Alice Wallenberg Foundation, Kavli Foundation, and Istituto Nazionale di Fisica Nucleare. This project has received funding or support from the European Union’s Horizon 2020 research and innovation programme under the Marie Sklodowska-Curie Grant Agreements No. 690575 and No. 674896, respectively. Data processing is performed using infrastructures from the Open Science Grid and European Grid Initiative. We are grateful to Laboratori Nazionali del Gran Sasso for hosting and supporting the XENON project.


\begin{thebibliography}{43}%
\makeatletter
\providecommand \@ifxundefined [1]{%
 \@ifx{#1\undefined}
}%
\providecommand \@ifnum [1]{%
 \ifnum #1\expandafter \@firstoftwo
 \else \expandafter \@secondoftwo
 \fi
}%
\providecommand \@ifx [1]{%
 \ifx #1\expandafter \@firstoftwo
 \else \expandafter \@secondoftwo
 \fi
}%
\providecommand \natexlab [1]{#1}%
\providecommand \enquote  [1]{``#1''}%
\providecommand \bibnamefont  [1]{#1}%
\providecommand \bibfnamefont [1]{#1}%
\providecommand \citenamefont [1]{#1}%
\providecommand \href@noop [0]{\@secondoftwo}%
\providecommand \href [0]{\begingroup \@sanitize@url \@href}%
\providecommand \@href[1]{\@@startlink{#1}\@@href}%
\providecommand \@@href[1]{\endgroup#1\@@endlink}%
\providecommand \@sanitize@url [0]{\catcode `\\12\catcode `\$12\catcode
  `\&12\catcode `\#12\catcode `\^12\catcode `\_12\catcode `\%12\relax}%
\providecommand \@@startlink[1]{}%
\providecommand \@@endlink[0]{}%
\providecommand \url  [0]{\begingroup\@sanitize@url \@url }%
\providecommand \@url [1]{\endgroup\@href {#1}{\urlprefix }}%
\providecommand \urlprefix  [0]{URL }%
\providecommand \Eprint [0]{\href }%
\providecommand \doibase [0]{http://dx.doi.org/}%
\providecommand \selectlanguage [0]{\@gobble}%
\providecommand \bibinfo  [0]{\@secondoftwo}%
\providecommand \bibfield  [0]{\@secondoftwo}%
\providecommand \translation [1]{[#1]}%
\providecommand \BibitemOpen [0]{}%
\providecommand \bibitemStop [0]{}%
\providecommand \bibitemNoStop [0]{.\EOS\space}%
\providecommand \EOS [0]{\spacefactor3000\relax}%
\providecommand \BibitemShut  [1]{\csname bibitem#1\endcsname}%
\let\auto@bib@innerbib\@empty
%</preamble>
\bibitem [{\citenamefont {Clowe}\ \emph {et~al.}(2004)\citenamefont {Clowe},
  \citenamefont {Gonzalez},\ and\ \citenamefont {Markevitch}}]{clowe2004weak}%
  \BibitemOpen
  \bibfield  {author} {\bibinfo {author} {\bibfnamefont {D.}~\bibnamefont
  {Clowe}}, \bibinfo {author} {\bibfnamefont {A.}~\bibnamefont {Gonzalez}}, \
  and\ \bibinfo {author} {\bibfnamefont {M.}~\bibnamefont {Markevitch}},\
  }\href@noop {} {\bibfield  {journal} {\bibinfo  {journal} {ApJ}\ }\textbf
  {\bibinfo {volume} {604}},\ \bibinfo {pages} {596} (\bibinfo {year}
  {2004})}\BibitemShut {NoStop}%
\bibitem [{\citenamefont {Rubin}\ \emph {et~al.}(1980)\citenamefont {Rubin},
  \citenamefont {Ford~Jr},\ and\ \citenamefont
  {Thonnard}}]{rubin1980rotational}%
  \BibitemOpen
  \bibfield  {author} {\bibinfo {author} {\bibfnamefont {V.~C.}\ \bibnamefont
  {Rubin}}, \bibinfo {author} {\bibfnamefont {W.~K.}\ \bibnamefont {Ford~Jr}},
  \ and\ \bibinfo {author} {\bibfnamefont {N.}~\bibnamefont {Thonnard}},\
  }\href@noop {} {\bibfield  {journal} {\bibinfo  {journal} {ApJ}\ }\textbf
  {\bibinfo {volume} {238}},\ \bibinfo {pages} {471} (\bibinfo {year}
  {1980})}\BibitemShut {NoStop}%
\bibitem [{\citenamefont {Aghanim}\ \emph {et~al.}(2018)\citenamefont {Aghanim}
  \emph {et~al.}}]{aghanim2018planck}%
  \BibitemOpen
  \bibfield  {author} {\bibinfo {author} {\bibfnamefont {N.}~\bibnamefont
  {Aghanim}} \emph {et~al.} (\bibinfo {collaboration} {Planck}),\ }\href@noop
  {} {\bibfield  {journal} {\bibinfo  {journal} {arXiv:1807.06209}\ } (\bibinfo
  {year} {2018})}\BibitemShut {NoStop}%
\bibitem [{\citenamefont {Jungman}\ \emph {et~al.}(1996)\citenamefont
  {Jungman}, \citenamefont {Kamionkowski},\ and\ \citenamefont
  {Griest}}]{wimp_theory1}%
  \BibitemOpen
  \bibfield  {author} {\bibinfo {author} {\bibfnamefont {G.}~\bibnamefont
  {Jungman}}, \bibinfo {author} {\bibfnamefont {M.}~\bibnamefont
  {Kamionkowski}}, \ and\ \bibinfo {author} {\bibfnamefont {K.}~\bibnamefont
  {Griest}},\ }\href@noop {} {\bibfield  {journal} {\bibinfo  {journal} {Phys.
  Rept.}\ }\textbf {\bibinfo {volume} {267}},\ \bibinfo {pages} {195} (\bibinfo
  {year} {1996})}\BibitemShut {NoStop}%
\bibitem [{\citenamefont {Lee}\ and\ \citenamefont
  {Weinberg}(1977)}]{thermal_relic}%
  \BibitemOpen
  \bibfield  {author} {\bibinfo {author} {\bibfnamefont {B.~W.}\ \bibnamefont
  {Lee}}\ and\ \bibinfo {author} {\bibfnamefont {S.}~\bibnamefont {Weinberg}},\
  }\href@noop {} {\bibfield  {journal} {\bibinfo  {journal} {Phys. Rev. Lett.}\
  }\textbf {\bibinfo {volume} {39}},\ \bibinfo {pages} {165} (\bibinfo {year}
  {1977})}\BibitemShut {NoStop}%
\bibitem [{\citenamefont {Akerib}\ \emph {et~al.}(2016)\citenamefont {Akerib}
  \emph {et~al.}}]{lux}%
  \BibitemOpen
  \bibfield  {author} {\bibinfo {author} {\bibfnamefont {D.~S.}\ \bibnamefont
  {Akerib}} \emph {et~al.} (\bibinfo {collaboration} {LUX}),\ }\href@noop {}
  {\bibfield  {journal} {\bibinfo  {journal} {Phys. Rev. Lett.}\ }\textbf
  {\bibinfo {volume} {116}},\ \bibinfo {pages} {161302} (\bibinfo {year}
  {2016})}\BibitemShut {NoStop}%
\bibitem [{\citenamefont {Cui}\ \emph {et~al.}(2017)\citenamefont {Cui} \emph
  {et~al.}}]{pandax}%
  \BibitemOpen
  \bibfield  {author} {\bibinfo {author} {\bibfnamefont {X.}~\bibnamefont
  {Cui}} \emph {et~al.} (\bibinfo {collaboration} {PandaX-II}),\ }\href@noop {}
  {\bibfield  {journal} {\bibinfo  {journal} {Phys. Rev. Lett.}\ }\textbf
  {\bibinfo {volume} {119}},\ \bibinfo {pages} {181302} (\bibinfo {year}
  {2017})}\BibitemShut {NoStop}%
\bibitem [{\citenamefont {Aprile}\ \emph {et~al.}(2018)\citenamefont {Aprile}
  \emph {et~al.}}]{xe1t_sr1}%
  \BibitemOpen
  \bibfield  {author} {\bibinfo {author} {\bibfnamefont {E.}~\bibnamefont
  {Aprile}} \emph {et~al.} (\bibinfo {collaboration} {XENON}),\ }\href@noop {}
  {\bibfield  {journal} {\bibinfo  {journal} {Phys. Rev. Lett.}\ }\textbf
  {\bibinfo {volume} {121}},\ \bibinfo {pages} {111302} (\bibinfo {year}
  {2018})}\BibitemShut {NoStop}%
\bibitem [{\citenamefont {Battaglieri}\ \emph {et~al.}(2017)\citenamefont
  {Battaglieri} \emph {et~al.}}]{low_mass_dm}%
  \BibitemOpen
  \bibfield  {author} {\bibinfo {author} {\bibfnamefont {M.}~\bibnamefont
  {Battaglieri}} \emph {et~al.},\ }\href@noop {} {\bibfield  {journal}
  {\bibinfo  {journal} {arXiv:1707.04591}\ } (\bibinfo {year}
  {2017})}\BibitemShut {NoStop}%
\bibitem [{\citenamefont {Boehm}\ \emph {et~al.}(2004)\citenamefont {Boehm},
  \citenamefont {Ensslin},\ and\ \citenamefont {Silk}}]{boehm2004can}%
  \BibitemOpen
  \bibfield  {author} {\bibinfo {author} {\bibfnamefont {C.}~\bibnamefont
  {Boehm}}, \bibinfo {author} {\bibfnamefont {T.}~\bibnamefont {Ensslin}}, \
  and\ \bibinfo {author} {\bibfnamefont {J.}~\bibnamefont {Silk}},\ }\href@noop
  {} {\bibfield  {journal} {\bibinfo  {journal} {Journal of Physics G: Nuclear
  and Particle Physics}\ }\textbf {\bibinfo {volume} {30}},\ \bibinfo {pages}
  {279} (\bibinfo {year} {2004})}\BibitemShut {NoStop}%
\bibitem [{\citenamefont {Boehm}\ and\ \citenamefont
  {Fayet}(2004)}]{boehm2004scalar}%
  \BibitemOpen
  \bibfield  {author} {\bibinfo {author} {\bibfnamefont {C.}~\bibnamefont
  {Boehm}}\ and\ \bibinfo {author} {\bibfnamefont {P.}~\bibnamefont {Fayet}},\
  }\href@noop {} {\bibfield  {journal} {\bibinfo  {journal} {Nuclear Physics
  B}\ }\textbf {\bibinfo {volume} {683}},\ \bibinfo {pages} {219} (\bibinfo
  {year} {2004})}\BibitemShut {NoStop}%
\bibitem [{\citenamefont {Kouvaris}\ and\ \citenamefont
  {Pradler}(2017)}]{bremsstrahlung}%
  \BibitemOpen
  \bibfield  {author} {\bibinfo {author} {\bibfnamefont {C.}~\bibnamefont
  {Kouvaris}}\ and\ \bibinfo {author} {\bibfnamefont {J.}~\bibnamefont
  {Pradler}},\ }\href@noop {} {\bibfield  {journal} {\bibinfo  {journal} {Phys.
  Rev. Lett.}\ }\textbf {\bibinfo {volume} {118}},\ \bibinfo {pages} {031803}
  (\bibinfo {year} {2017})}\BibitemShut {NoStop}%
\bibitem [{\citenamefont {Migdal}(1941)}]{original_migdal}%
  \BibitemOpen
  \bibfield  {author} {\bibinfo {author} {\bibfnamefont {A.}~\bibnamefont
  {Migdal}},\ }\href@noop {} {\bibfield  {journal} {\bibinfo  {journal} {J.
  Phys.(USSR)}\ }\textbf {\bibinfo {volume} {4}},\ \bibinfo {pages} {449}
  (\bibinfo {year} {1941})}\BibitemShut {NoStop}%
\bibitem [{\citenamefont {Ibe}\ \emph {et~al.}(2018)\citenamefont {Ibe},
  \citenamefont {Nakano}, \citenamefont {Shoji},\ and\ \citenamefont
  {Suzuki}}]{migdal}%
  \BibitemOpen
  \bibfield  {author} {\bibinfo {author} {\bibfnamefont {M.}~\bibnamefont
  {Ibe}}, \bibinfo {author} {\bibfnamefont {W.}~\bibnamefont {Nakano}},
  \bibinfo {author} {\bibfnamefont {Y.}~\bibnamefont {Shoji}}, \ and\ \bibinfo
  {author} {\bibfnamefont {K.}~\bibnamefont {Suzuki}},\ }\href@noop {}
  {\bibfield  {journal} {\bibinfo  {journal} {JHEP}\ }\textbf {\bibinfo
  {volume} {03}},\ \bibinfo {pages} {194} (\bibinfo {year} {2018})}\BibitemShut
  {NoStop}%
\bibitem [{\citenamefont {Aprile}\ \emph
  {et~al.}(2017{\natexlab{a}})\citenamefont {Aprile} \emph
  {et~al.}}]{xe1t_instrument}%
  \BibitemOpen
  \bibfield  {author} {\bibinfo {author} {\bibfnamefont {E.}~\bibnamefont
  {Aprile}} \emph {et~al.} (\bibinfo {collaboration} {XENON}),\ }\href@noop {}
  {\bibfield  {journal} {\bibinfo  {journal} {Eur. Phys. J. C}\ }\textbf
  {\bibinfo {volume} {77}},\ \bibinfo {pages} {881} (\bibinfo {year}
  {2017}{\natexlab{a}})}\BibitemShut {NoStop}%
\bibitem [{\citenamefont {Aprile}\ \emph {et~al.}(2015)\citenamefont {Aprile}
  \emph {et~al.}}]{pmt1}%
  \BibitemOpen
  \bibfield  {author} {\bibinfo {author} {\bibfnamefont {E.}~\bibnamefont
  {Aprile}} \emph {et~al.} (\bibinfo {collaboration} {XENON}),\ }\href@noop {}
  {\bibfield  {journal} {\bibinfo  {journal} {Eur. Phys. J. C}\ }\textbf
  {\bibinfo {volume} {75}},\ \bibinfo {pages} {546} (\bibinfo {year}
  {2015})}\BibitemShut {NoStop}%
\bibitem [{\citenamefont {Aprile}\ \emph
  {et~al.}(2017{\natexlab{b}})\citenamefont {Aprile} \emph {et~al.}}]{pmt2}%
  \BibitemOpen
  \bibfield  {author} {\bibinfo {author} {\bibfnamefont {E.}~\bibnamefont
  {Aprile}} \emph {et~al.} (\bibinfo {collaboration} {XENON}),\ }\href@noop {}
  {\bibfield  {journal} {\bibinfo  {journal} {JINST}\ }\textbf {\bibinfo
  {volume} {12}},\ \bibinfo {pages} {no. 01, P01024} (\bibinfo {year}
  {2017}{\natexlab{b}})}\BibitemShut {NoStop}%
\bibitem [{\citenamefont {Aprile}\ \emph
  {et~al.}(2019{\natexlab{a}})\citenamefont {Aprile} \emph
  {et~al.}}]{xe1t_analysis1}%
  \BibitemOpen
  \bibfield  {author} {\bibinfo {author} {\bibfnamefont {E.}~\bibnamefont
  {Aprile}} \emph {et~al.} (\bibinfo {collaboration} {XENON}),\ }\href@noop {}
  {\bibfield  {journal} {\bibinfo  {journal} {Phys. Rev. D}\ }\textbf {\bibinfo
  {volume} {100}},\ \bibinfo {pages} {052014} (\bibinfo {year}
  {2019}{\natexlab{a}})}\BibitemShut {NoStop}%
\bibitem [{\citenamefont {Aprile}\ \emph
  {et~al.}(2019{\natexlab{b}})\citenamefont {Aprile} \emph
  {et~al.}}]{xe1t_analysis2}%
  \BibitemOpen
  \bibfield  {author} {\bibinfo {author} {\bibfnamefont {E.}~\bibnamefont
  {Aprile}} \emph {et~al.} (\bibinfo {collaboration} {XENON}),\ }\href@noop {}
  {\bibfield  {journal} {\bibinfo  {journal} {Phys. Rev. D}\ }\textbf {\bibinfo
  {volume} {99}},\ \bibinfo {pages} {112009} (\bibinfo {year}
  {2019}{\natexlab{b}})}\BibitemShut {NoStop}%
\bibitem [{\citenamefont {Lindhard}\ \emph {et~al.}(1963)\citenamefont
  {Lindhard}, \citenamefont {Nielsen}, \citenamefont {Scharff},\ and\
  \citenamefont {Thomsen}}]{lindhard1963integral}%
  \BibitemOpen
  \bibfield  {author} {\bibinfo {author} {\bibfnamefont {J.}~\bibnamefont
  {Lindhard}}, \bibinfo {author} {\bibfnamefont {V.}~\bibnamefont {Nielsen}},
  \bibinfo {author} {\bibfnamefont {M.}~\bibnamefont {Scharff}}, \ and\
  \bibinfo {author} {\bibfnamefont {P.}~\bibnamefont {Thomsen}},\ }\href@noop
  {} {\bibfield  {journal} {\bibinfo  {journal} {Kgl. Danske Videnskab.,
  Selskab. Mat. Fys. Medd.}\ }\textbf {\bibinfo {volume} {33}} (\bibinfo {year}
  {1963})}\BibitemShut {NoStop}%
\bibitem [{\citenamefont {Akerib}\ \emph {et~al.}(2017)\citenamefont {Akerib}
  \emph {et~al.}}]{xe127_lux}%
  \BibitemOpen
  \bibfield  {author} {\bibinfo {author} {\bibfnamefont {D.~S.}\ \bibnamefont
  {Akerib}} \emph {et~al.} (\bibinfo {collaboration} {LUX}),\ }\href@noop {}
  {\bibfield  {journal} {\bibinfo  {journal} {Phys. Rev. D}\ }\textbf {\bibinfo
  {volume} {96}},\ \bibinfo {pages} {112011} (\bibinfo {year}
  {2017})}\BibitemShut {NoStop}%
\bibitem [{\citenamefont {Boulton}\ \emph {et~al.}(2017)\citenamefont
  {Boulton}, \citenamefont {Bernard}, \citenamefont {Destefano}, \citenamefont
  {Edwards}, \citenamefont {Gai}, \citenamefont {Hertel}, \citenamefont {Horn},
  \citenamefont {Larsen}, \citenamefont {Tennyson}, \citenamefont {Wahl} \emph
  {et~al.}}]{boulton2017calibration}%
  \BibitemOpen
  \bibfield  {author} {\bibinfo {author} {\bibfnamefont {E.}~\bibnamefont
  {Boulton}}, \bibinfo {author} {\bibfnamefont {E.}~\bibnamefont {Bernard}},
  \bibinfo {author} {\bibfnamefont {N.}~\bibnamefont {Destefano}}, \bibinfo
  {author} {\bibfnamefont {B.}~\bibnamefont {Edwards}}, \bibinfo {author}
  {\bibfnamefont {M.}~\bibnamefont {Gai}}, \bibinfo {author} {\bibfnamefont
  {S.}~\bibnamefont {Hertel}}, \bibinfo {author} {\bibfnamefont
  {M.}~\bibnamefont {Horn}}, \bibinfo {author} {\bibfnamefont {N.}~\bibnamefont
  {Larsen}}, \bibinfo {author} {\bibfnamefont {B.}~\bibnamefont {Tennyson}},
  \bibinfo {author} {\bibfnamefont {C.}~\bibnamefont {Wahl}},  \emph {et~al.},\
  }\href@noop {} {\bibfield  {journal} {\bibinfo  {journal} {Journal of
  Instrumentation}\ }\textbf {\bibinfo {volume} {12}},\ \bibinfo {pages}
  {P08004} (\bibinfo {year} {2017})}\BibitemShut {NoStop}%
\bibitem [{\citenamefont {Aprile}\ \emph
  {et~al.}(2019{\natexlab{c}})\citenamefont {Aprile} \emph
  {et~al.}}]{xe1t_s2only}%
  \BibitemOpen
  \bibfield  {author} {\bibinfo {author} {\bibfnamefont {E.}~\bibnamefont
  {Aprile}} \emph {et~al.} (\bibinfo {collaboration} {XENON}),\ }\href@noop {}
  {\bibfield  {journal} {\bibinfo  {journal} {arXiv: 1907.11485}\ } (\bibinfo
  {year} {2019}{\natexlab{c}})}\BibitemShut {NoStop}%
\bibitem [{\citenamefont {Szydagis}\ \emph {et~al.}(2018)\citenamefont
  {Szydagis} \emph {et~al.}}]{nest_v2}%
  \BibitemOpen
  \bibfield  {author} {\bibinfo {author} {\bibfnamefont {M.}~\bibnamefont
  {Szydagis}} \emph {et~al.},\ }\href {\doibase 10.5281/zenodo.1314669}
  {\enquote {\bibinfo {title} {Noble element simulation technique v2.0},}\ }
  (\bibinfo {year} {2018})\BibitemShut {NoStop}%
\bibitem [{\citenamefont {Akerib}\ \emph {et~al.}(2019)\citenamefont {Akerib}
  \emph {et~al.}}]{lux_migdal}%
  \BibitemOpen
  \bibfield  {author} {\bibinfo {author} {\bibfnamefont {D.~S.}\ \bibnamefont
  {Akerib}} \emph {et~al.} (\bibinfo {collaboration} {LUX}),\ }\href@noop {}
  {\bibfield  {journal} {\bibinfo  {journal} {Phys. Rev. Lett.}\ }\textbf
  {\bibinfo {volume} {122}},\ \bibinfo {pages} {131301} (\bibinfo {year}
  {2019})}\BibitemShut {NoStop}%
\bibitem [{\citenamefont {Armengaud}\ \emph {et~al.}(2019)\citenamefont
  {Armengaud} \emph {et~al.}}]{edelweiss}%
  \BibitemOpen
  \bibfield  {author} {\bibinfo {author} {\bibfnamefont {E.}~\bibnamefont
  {Armengaud}} \emph {et~al.} (\bibinfo {collaboration} {EDELWEISS}),\
  }\href@noop {} {\bibfield  {journal} {\bibinfo  {journal} {Phys. Rev. D}\
  }\textbf {\bibinfo {volume} {99}},\ \bibinfo {pages} {082003} (\bibinfo
  {year} {2019})}\BibitemShut {NoStop}%
\bibitem [{\citenamefont {Liu}\ \emph {et~al.}(2019)\citenamefont {Liu} \emph
  {et~al.}}]{cdex}%
  \BibitemOpen
  \bibfield  {author} {\bibinfo {author} {\bibfnamefont {Z.}~\bibnamefont
  {Liu}} \emph {et~al.},\ }\href@noop {} {\bibfield  {journal} {\bibinfo
  {journal} {Phys. Rev. Lett.}\ }\textbf {\bibinfo {volume} {123}},\ \bibinfo
  {pages} {161301} (\bibinfo {year} {2019})}\BibitemShut {NoStop}%
\bibitem [{\citenamefont {Abdelhameed}\ \emph {et~al.}(2017)\citenamefont
  {Abdelhameed} \emph {et~al.}}]{cresst_iii}%
  \BibitemOpen
  \bibfield  {author} {\bibinfo {author} {\bibfnamefont {A.}~\bibnamefont
  {Abdelhameed}} \emph {et~al.} (\bibinfo {collaboration} {CRESST}),\
  }\href@noop {} {\bibfield  {journal} {\bibinfo  {journal} {arXiv:1904.00498}\
  } (\bibinfo {year} {2017})}\BibitemShut {NoStop}%
\bibitem [{\citenamefont {Arnaud}\ \emph {et~al.}(2018)\citenamefont {Arnaud}
  \emph {et~al.}}]{news_g}%
  \BibitemOpen
  \bibfield  {author} {\bibinfo {author} {\bibfnamefont {Q.}~\bibnamefont
  {Arnaud}} \emph {et~al.} (\bibinfo {collaboration} {NEWS-G}),\ }\href@noop {}
  {\bibfield  {journal} {\bibinfo  {journal} {Astropart. Phys.}\ }\textbf
  {\bibinfo {volume} {97}},\ \bibinfo {pages} {54} (\bibinfo {year}
  {2018})}\BibitemShut {NoStop}%
\bibitem [{\citenamefont {Agnese}\ \emph {et~al.}(2016)\citenamefont {Agnese}
  \emph {et~al.}}]{cdmslite_ii}%
  \BibitemOpen
  \bibfield  {author} {\bibinfo {author} {\bibfnamefont {R.}~\bibnamefont
  {Agnese}} \emph {et~al.} (\bibinfo {collaboration} {SuperCDMS}),\ }\href@noop
  {} {\bibfield  {journal} {\bibinfo  {journal} {Phys. Rev. Lett.}\ }\textbf
  {\bibinfo {volume} {116}},\ \bibinfo {pages} {071301} (\bibinfo {year}
  {2016})}\BibitemShut {NoStop}%
\bibitem [{\citenamefont {Agnes}\ \emph {et~al.}(2018)\citenamefont {Agnes}
  \emph {et~al.}}]{darkside_s2only}%
  \BibitemOpen
  \bibfield  {author} {\bibinfo {author} {\bibfnamefont {P.}~\bibnamefont
  {Agnes}} \emph {et~al.} (\bibinfo {collaboration} {DarkSide}),\ }\href@noop
  {} {\bibfield  {journal} {\bibinfo  {journal} {Phys. Rev. Lett.}\ }\textbf
  {\bibinfo {volume} {121}},\ \bibinfo {pages} {081307} (\bibinfo {year}
  {2018})}\BibitemShut {NoStop}%
\bibitem [{\citenamefont {Abdelhameed}\ \emph {et~al.}(2019)\citenamefont
  {Abdelhameed} \emph {et~al.}}]{cresst_sd}%
  \BibitemOpen
  \bibfield  {author} {\bibinfo {author} {\bibfnamefont {A.}~\bibnamefont
  {Abdelhameed}} \emph {et~al.} (\bibinfo {collaboration} {CRESST}),\
  }\href@noop {} {\bibfield  {journal} {\bibinfo  {journal} {Eur. Phys. J. C}\
  }\textbf {\bibinfo {volume} {79}},\ \bibinfo {pages} {630} (\bibinfo {year}
  {2019})}\BibitemShut {NoStop}%
\bibitem [{\citenamefont {Agnese}\ \emph {et~al.}(2018)\citenamefont {Agnese}
  \emph {et~al.}}]{cdmslite_sd}%
  \BibitemOpen
  \bibfield  {author} {\bibinfo {author} {\bibfnamefont {R.}~\bibnamefont
  {Agnese}} \emph {et~al.} (\bibinfo {collaboration} {CDMS}),\ }\href@noop {}
  {\bibfield  {journal} {\bibinfo  {journal} {Phys. Rev. D}\ }\textbf {\bibinfo
  {volume} {97}},\ \bibinfo {pages} {022002} (\bibinfo {year}
  {2018})}\BibitemShut {NoStop}%
\bibitem [{\citenamefont {Lewin}\ and\ \citenamefont {Smith}(1996)}]{astro_dm}%
  \BibitemOpen
  \bibfield  {author} {\bibinfo {author} {\bibfnamefont {J.}~\bibnamefont
  {Lewin}}\ and\ \bibinfo {author} {\bibfnamefont {P.}~\bibnamefont {Smith}},\
  }\href@noop {} {\bibfield  {journal} {\bibinfo  {journal} {Astropart. Phys.}\
  }\textbf {\bibinfo {volume} {6}},\ \bibinfo {pages} {87} (\bibinfo {year}
  {1996})}\BibitemShut {NoStop}%
\bibitem [{\citenamefont {Aprile}\ \emph
  {et~al.}(2019{\natexlab{d}})\citenamefont {Aprile} \emph {et~al.}}]{xe1t_sd}%
  \BibitemOpen
  \bibfield  {author} {\bibinfo {author} {\bibfnamefont {E.}~\bibnamefont
  {Aprile}} \emph {et~al.} (\bibinfo {collaboration} {XENON}),\ }\href@noop {}
  {\bibfield  {journal} {\bibinfo  {journal} {Phys. Rev. Lett.}\ }\textbf
  {\bibinfo {volume} {122}},\ \bibinfo {pages} {141301} (\bibinfo {year}
  {2019}{\natexlab{d}})}\BibitemShut {NoStop}%
\bibitem [{\citenamefont {Del~Nobile}\ \emph {et~al.}(2015)\citenamefont
  {Del~Nobile}, \citenamefont {Kaplinghat},\ and\ \citenamefont
  {Yu}}]{lightmediator}%
  \BibitemOpen
  \bibfield  {author} {\bibinfo {author} {\bibfnamefont {E.}~\bibnamefont
  {Del~Nobile}}, \bibinfo {author} {\bibfnamefont {M.}~\bibnamefont
  {Kaplinghat}}, \ and\ \bibinfo {author} {\bibfnamefont {H.~B.}\ \bibnamefont
  {Yu}},\ }\href@noop {} {\bibfield  {journal} {\bibinfo  {journal} {JCAP}\
  }\textbf {\bibinfo {volume} {10}},\ \bibinfo {pages} {055} (\bibinfo {year}
  {2015})}\BibitemShut {NoStop}%
\bibitem [{\citenamefont {Ren}\ \emph {et~al.}(2018)\citenamefont {Ren} \emph
  {et~al.}}]{pandax_lm}%
  \BibitemOpen
  \bibfield  {author} {\bibinfo {author} {\bibfnamefont {X.}~\bibnamefont
  {Ren}} \emph {et~al.} (\bibinfo {collaboration} {PandaX-II}),\ }\href@noop {}
  {\bibfield  {journal} {\bibinfo  {journal} {Phys. Rev. Lett.}\ }\textbf
  {\bibinfo {volume} {121}},\ \bibinfo {pages} {021304} (\bibinfo {year}
  {2018})}\BibitemShut {NoStop}%
\bibitem [{\citenamefont {Emken}\ and\ \citenamefont
  {Kouvaris}(2018)}]{earth_shielding_ref1}%
  \BibitemOpen
  \bibfield  {author} {\bibinfo {author} {\bibfnamefont {T.}~\bibnamefont
  {Emken}}\ and\ \bibinfo {author} {\bibfnamefont {C.}~\bibnamefont
  {Kouvaris}},\ }\href@noop {} {\bibfield  {journal} {\bibinfo  {journal}
  {Phys. Rev. D}\ }\textbf {\bibinfo {volume} {97}},\ \bibinfo {pages} {115047}
  (\bibinfo {year} {2018})}\BibitemShut {NoStop}%
\bibitem [{\citenamefont {Mahdawi}\ and\ \citenamefont
  {Farrar}(2017)}]{earth_shielding_ref2}%
  \BibitemOpen
  \bibfield  {author} {\bibinfo {author} {\bibfnamefont {M.~S.}\ \bibnamefont
  {Mahdawi}}\ and\ \bibinfo {author} {\bibfnamefont {G.~R.}\ \bibnamefont
  {Farrar}},\ }\href@noop {} {\bibfield  {journal} {\bibinfo  {journal}
  {arXiv:1712.01170}\ } (\bibinfo {year} {2017})}\BibitemShut {NoStop}%
\bibitem [{\citenamefont {Emken}\ and\ \citenamefont
  {Kouvaris}(2017)}]{earth_shielding_ref3}%
  \BibitemOpen
  \bibfield  {author} {\bibinfo {author} {\bibfnamefont {T.}~\bibnamefont
  {Emken}}\ and\ \bibinfo {author} {\bibfnamefont {C.}~\bibnamefont
  {Kouvaris}},\ }\href@noop {} {\bibfield  {journal} {\bibinfo  {journal}
  {JCAP}\ }\textbf {\bibinfo {volume} {10}},\ \bibinfo {pages} {031} (\bibinfo
  {year} {2017})}\BibitemShut {NoStop}%
\bibitem [{\citenamefont {Kavanagh}(2017)}]{verne}%
  \BibitemOpen
  \bibfield  {author} {\bibinfo {author} {\bibfnamefont {B.~J.}\ \bibnamefont
  {Kavanagh}},\ }\href {\doibase 10.5281/zenodo.1116305} {\enquote {\bibinfo
  {title} {bradkav/verne: Release},}\ } (\bibinfo {year} {2017})\BibitemShut
  {NoStop}%
\bibitem [{\citenamefont {Kavanagh}(2018)}]{verne_method}%
  \BibitemOpen
  \bibfield  {author} {\bibinfo {author} {\bibfnamefont {B.~J.}\ \bibnamefont
  {Kavanagh}},\ }\href@noop {} {\bibfield  {journal} {\bibinfo  {journal}
  {Phys. Rev. D}\ }\textbf {\bibinfo {volume} {97}},\ \bibinfo {pages} {123013}
  (\bibinfo {year} {2018})}\BibitemShut {NoStop}%
\bibitem [{\citenamefont {Hooper}\ and\ \citenamefont
  {McDermott}(2018)}]{sd_earth_shielding_ref}%
  \BibitemOpen
  \bibfield  {author} {\bibinfo {author} {\bibfnamefont {D.}~\bibnamefont
  {Hooper}}\ and\ \bibinfo {author} {\bibfnamefont {S.~D.}\ \bibnamefont
  {McDermott}},\ }\href@noop {} {\bibfield  {journal} {\bibinfo  {journal}
  {Phy. Rev. D}\ }\textbf {\bibinfo {volume} {97}},\ \bibinfo {pages} {115006}
  (\bibinfo {year} {2018})}\BibitemShut {NoStop}%
\end{thebibliography}
\end{document}